\documentclass[aps,twocolumn,prl,preprintnumbers,amsmath,amssymb,superscriptaddress]{revtex4}

\usepackage{graphicx}
\usepackage{bm}
\usepackage{hyperref}
\usepackage{natbib}

\begin{document}

\title{A hierarchy of bound states in the 1D ferromagnetic Ising chain CoNb$_2$O$_6$ investigated by high resolution time-domain terahertz spectroscopy}

\author{C. M. Morris}
\affiliation{The Institute for Quantum Matter, Department of Physics and Astronomy, The Johns Hopkins University, Baltimore, MD 21218, USA}

\author{R. Vald\'{e}s Aguilar}
\affiliation{The Institute for Quantum Matter, Department of Physics and Astronomy, The Johns Hopkins University, Baltimore, MD 21218, USA}
\affiliation{Center for Integrated Nanotechnologies, Los Alamos National Laboratory. MS K771. Los Alamos, NM 87545}

\author{A. Ghosh}
\affiliation{The Institute for Quantum Matter, Department of Physics and Astronomy, The Johns Hopkins University, Baltimore, MD 21218, USA}

\author{S. M. Koohpayeh}
\affiliation{The Institute for Quantum Matter, Department of Physics and Astronomy, The Johns Hopkins University, Baltimore, MD 21218, USA}

\author{J. Krizan}
\affiliation{Department of Chemistry, Princeton University, Princeton, NJ 08544, USA}

\author{R. J. Cava}
\affiliation{Department of Chemistry, Princeton University, Princeton, NJ 08544, USA}

 \author{O. Tchernyshyov}
\affiliation{The Institute for Quantum Matter, Department of Physics and Astronomy, The Johns Hopkins University, Baltimore, MD 21218, USA}

\author{T. M. McQueen}
\affiliation{The Institute for Quantum Matter, Department of Physics and Astronomy, The Johns Hopkins University, Baltimore, MD 21218, USA}
\affiliation{Department of Chemistry, The Johns Hopkins University, Baltimore, MD 21218, USA}

\author{N. P. Armitage}
\affiliation{The Institute for Quantum Matter, Department of Physics and Astronomy, The Johns Hopkins University, Baltimore, MD 21218, USA}

\date{\today}

\begin{abstract}

Kink bound states in the one dimensional ferromagnetic Ising chain compound CoNb$_2$O$_6$ have been studied using high resolution time-domain terahertz spectroscopy in zero applied magnetic field.  When magnetic order develops at low temperature, nine bound states of kinks become visible.  Their energies can be modeled exceedingly well by the Airy function solutions to a 1D Schr{\"o}dinger equation with a linear confining potential.  This sequence of bound states terminates at a threshold energy near two times the energy of the lowest bound state.  Above this energy scale we observe a broad feature consistent with the onset of the two particle continuum.  At energies just below this threshold we observe a prominent excitation that we interpret as a novel bound state of bound states -- two pairs of kinks on neighboring chains.
\end{abstract}

\maketitle

The one dimensional Ising spin chain is a paradigmatic example of an interacting quantum many body system.   Its low dimensionality increases its propensity for quantum fluctuations and hence its tendency to exhibit interesting quantum effects.   It has been proposed to host a number of exotic states of matter including ones with fractional excitations and novel quantum critical points (QCP) \cite{Levi2010,McCoy1978,Fogedby1978,Zamolodchikov1989}.  Moreover, the one-dimensionality often makes a theoretical formulation more tractable and a direct comparison between experiment and theory possible.

Recently, the Ising spin chain compound CoNb$_2$O$_6$  has been of interest due to a fascinating set of neutron scattering experiments performed by Coldea $et$ $al.$  \citep{Coldea2010}.  By applying a magnetic field transverse to the 1D ferromagnetic spin chains, they were able to tune through a QCP from a spin-ordered phase to a paramagnetic state.   At the QCP near 5.5 T they observed that the ratio of energies of the two lowest lying magnetic excitations was the golden ratio.   This was consistent with an emergent E$8$ symmetry at the transverse field-tuned QCP predicted by Zamolodchikov \cite{Zamolodchikov1989} for an Ising chain in the presence of a weak longitudinal field.

Ising spin chains also show interesting behavior in the absence of a transverse field when a weak longitudinal field is present.  At zero field the linear ferromagnetic Ising chain has a two-fold degenerate ground state. For an isolated chain, the excitations are sets of $n$ flipped spins, called ``spin clusters"\cite{Tinkham1969a,Tinkham1969b}. For pure Ising interactions, domains of different length are degenerate, as the exchange interaction is only broken at the ends of the cluster in domain walls or ``kinks"\cite{Pfeuty1970,Rutkevich08a}.  A longitudinal field lifts the degeneracy of domains of different length, with the energy to create longer chains increasing linearly with the separation between kinks.  Small interactions beyond pure Ising allow the kinks to move. \textcite{McCoy1978} solved for the excitations in this model, where two kinks can be treated as two particles moving in one dimension with a confining potential $\lambda |x|$ between them.  In the continuum approximation the discrete nature of the spin-chain can be ignored and the center-of-mass motion of the kinks can be described by a Schr{\"o}dinger equation

\begin{equation} \label{Sch}
-\frac{\hbar^2}{\mu}\frac{d^2}{d x^2} \psi(x)+ \lambda |x| \psi(x) = (m - 2m_0)\psi(x).
\end{equation}

Solutions to this equation are Airy functions with energy eigenvalues of the kink bound-states \cite{McCoy1978}

\begin{equation} \label{sps}
m_n = 2m_0 + z_n \lambda^{2/3} \left(\frac{\hbar^2}{\mu}\right)^{1/3}, \quad n=1,2,3\ldots
\end{equation}

\noindent where $m_n$ is the bound state energy, $m_0$ is the energy to break the nearest neighbor exchange interaction, $\lambda$ is proportional to the longitudinal magnetic field, and the $z_n$'s are the negative zeros of the Airy function of the first kind. In CoNb$_2$O$_6$, antiferromagnetic (AF) ordering of ferromagnetic chains below 1.97 K produces a weak effective longitudinal mean field.  In their neutron scattering measurements Coldea  $et$ $al.$  found a sequence of 5 bound states that correspond to the first 5 solutions of this Airy function model \citep{Coldea2010}. One can make an analogy between the linear confining potential and confinement in quantum chromodynamics, where a kink plays the role of a quark with bare mass $m_0$ and the kink bound state plays the role of a meson.  This analogy has been made recently for spin ladders \cite{Lake09a}.
 
In addition to neutron studies, insight into similar materials has been gained using far infrared light as a probe. Torrance and Tinkham  \cite{Tinkham1969a,Tinkham1969b} did pioneering work on one dimensional Ising chains, looking at CoCl$_2\cdot$2H$_2$O using a far-infrared grating spectrometer. They developed a model of the spin cluster excitations based on Ising basis functions, calculating the energies of discrete spin flip transitions in strong longitudinal field. In  CoNb$_2$O$_6$, some preliminary work was done using Fourier transform infrared spectroscopy and electron spin resonance \cite{Kunimoto1999a,Kunimoto1999b}.  However, the resolution was too low to resolve the excitations predicted by Eq. \ref{Sch}.

\begin{figure}[tb]
\includegraphics[width=1.0\columnwidth]{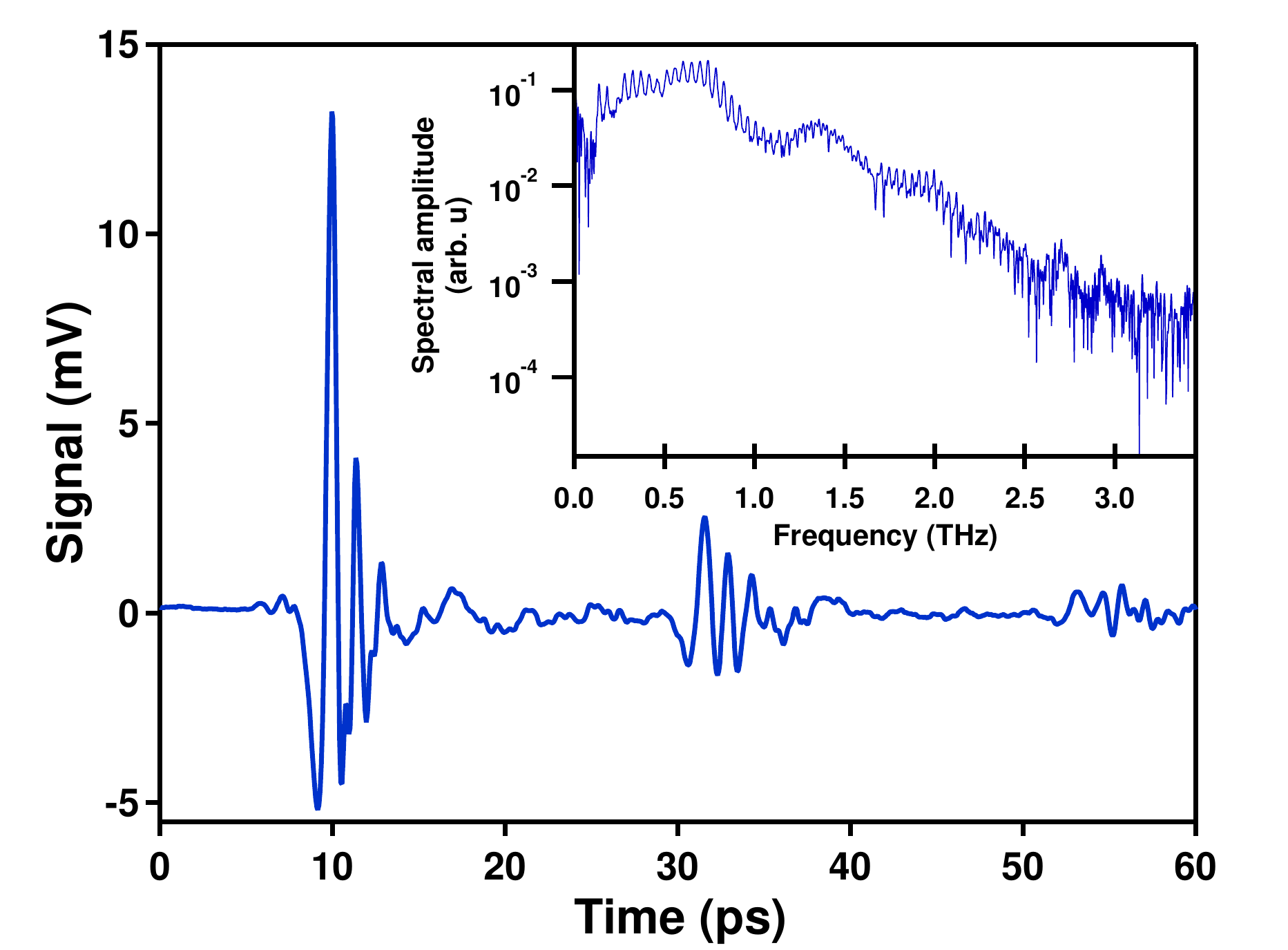}
\caption{Sample waveform at 5 K for $\mathbf{k}\|b$, $\mathbf{e}\|c$, $\mathbf{h}\|a$. Fine features in the spectral amplitude below 2.5 THz are not due to noise, but rather Fabry-Perot oscillations caused by the multiple reflections of the time domain pulse in the CoNb$_2$O$_6$ crystal.}
\label{Fig1}
\end{figure}

In this Letter we use high resolution time-domain terahertz spectroscopy (TDTS) to investigate these kink excitations in CoNb$_2$O$_6$ in the far infrared.   The high signal to noise, excellent energy resolution, and short acquisition times of the technique allow us to find a number of excitations that were previously unresolved by neutron scattering. In addition to the five kink bound states observed by  Coldea $et$ $al.$ \citep{Coldea2010}, terahertz spectroscopy shows a further four kink bound states and a new higher energy excitation below the continuum that we interpret as a bound state of bound states.

CoNb$_2$O$_6$ belongs to the orthorhombic $Pbcn$ space group.  Crystal field splitting produces an effective spin 1/2 moment on the Co$^{+2}$ ions, with the spins lying in the $ac$ plane at an angle of $\pm31^\circ$ to the $c-$axis \cite{Heid1995,Kobayashi2000}.  The Co atoms form zig-zag chains along the $c-$axis.  Ferromagnetic exchange interactions between nearest-neighbor Co$^{+2}$ ions along this axis cause ferromagnetic correlations in these chains beginning at $\sim$25 K \cite{Hanawa1994}.   Below 2.95 K weak AF inter-chain exchange interactions stabilize a spin-density wave along the $b$-direction with a temperature-dependent ordering wave vector $Q$.   Below 1.97 K the spin-density wave becomes commensurate AF along $b$ with a temperature independent $Q_{AF}=(0,1/2,0)$ and an ordered moment of $3.05 \mu_B$ \cite{Scharf79a}.   As mentioned above, in this low temperature phase the effects of weak interchain couplings can be understood as a small effective longitudinal field that scales with the ordered moment.  

The CoNb$_2$O$_6$ samples used here were grown by the floating zone method and characterized by powder and back-reflection X-ray Laue diffraction.   The samples were small discs approximately 5 mm in diameter and 600 $\mu$m thick.  Time-domain THz spectroscopy (TDTS) \cite{RevModPhys.83.543,Nuss1998} was performed using a home-built transmission mode spectrometer that can access the electrodynamic response between 100 GHz and 2 THz (0.41 meV -- 8.27 meV). Taking the ratio of the transmission through a sample to that of a reference aperture gives the complex transmission coefficient (See Supplementary Information (SI) for further details). The electric and magnetic fields of the terahertz waveform can be used to probe both electric and magnetic dipole excitations in the material. In the present case of magnetic insulators the time-varying magnetic field of the pulse couples to the spins of the system, essentially equivalent to frequency domain electron spin resonance (ESR).  As the wavelength of THz range radiation is much greater than typical lattice constants (1 THz $\sim$ 300 $\mu$m), TDTS measures the $q \rightarrow 0$ response.  The complex transmission $T(\omega)$ is related to the complex susceptibility $\chi(\omega)$ at $q=0$ as $-\ln(T(\omega)) \propto \omega \chi(q=0,\omega)$.

\begin{figure}[tb]
\includegraphics[width=0.9\columnwidth]{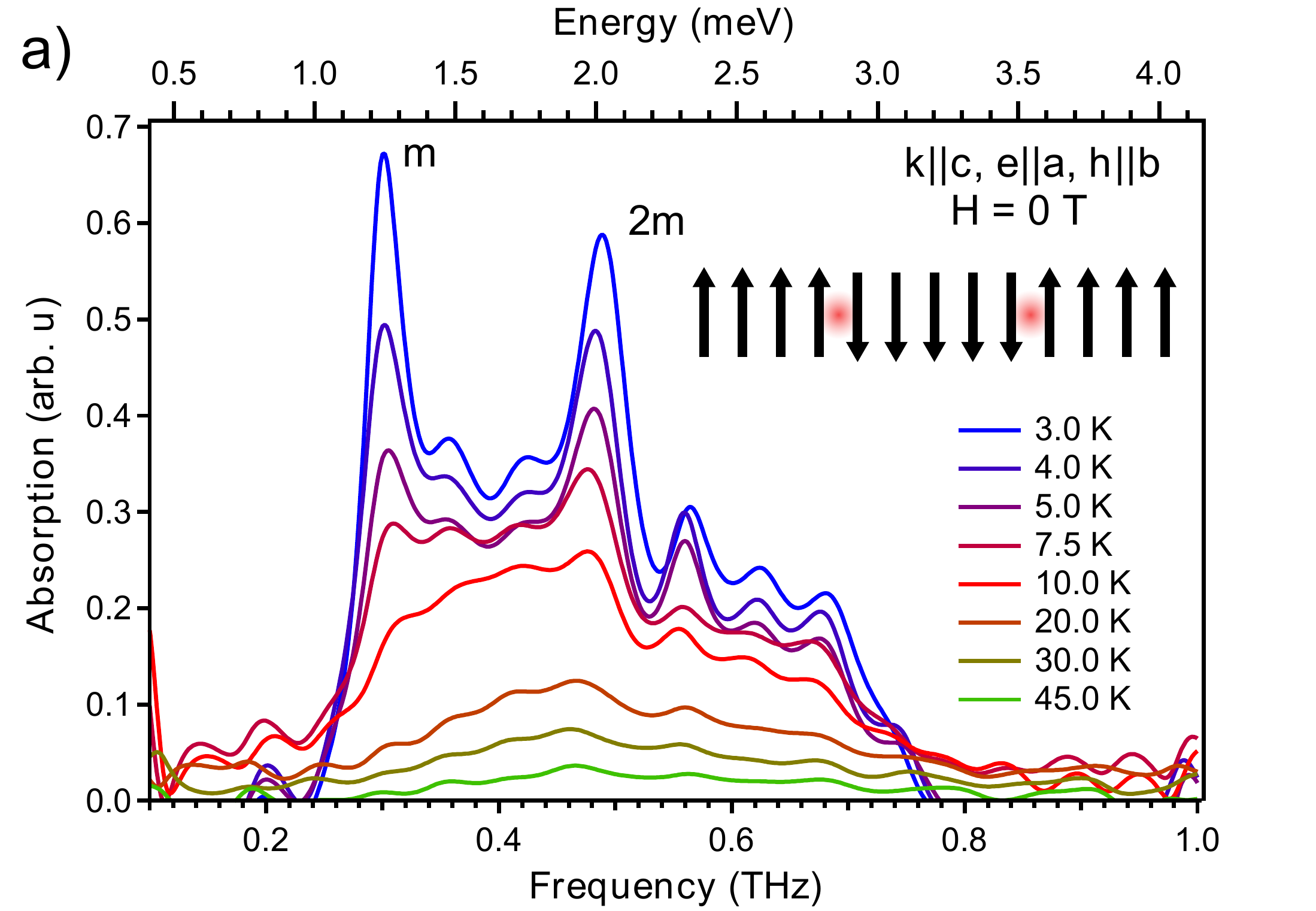}
\includegraphics[width=0.9\columnwidth]{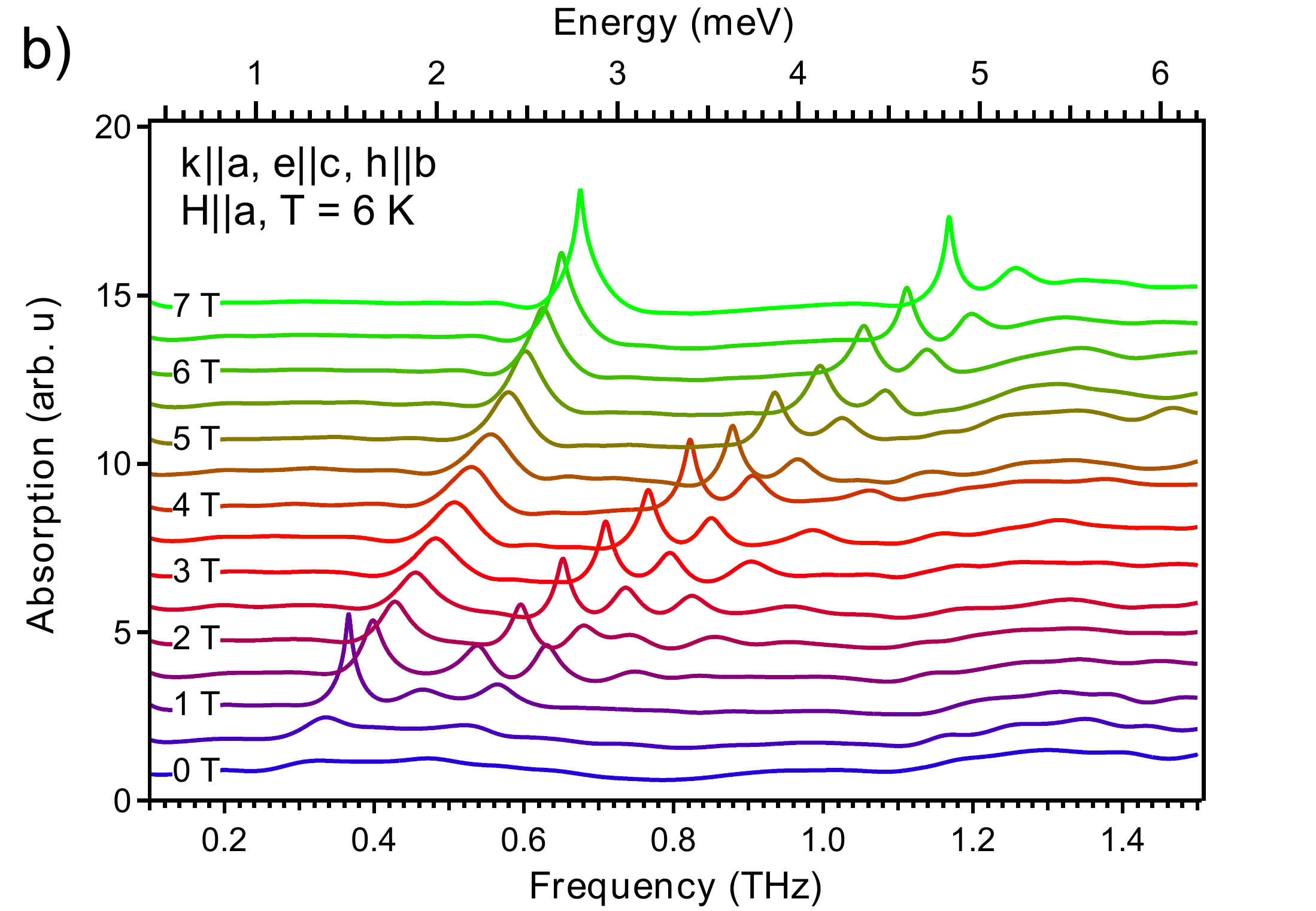}
\includegraphics[width=0.9\columnwidth]{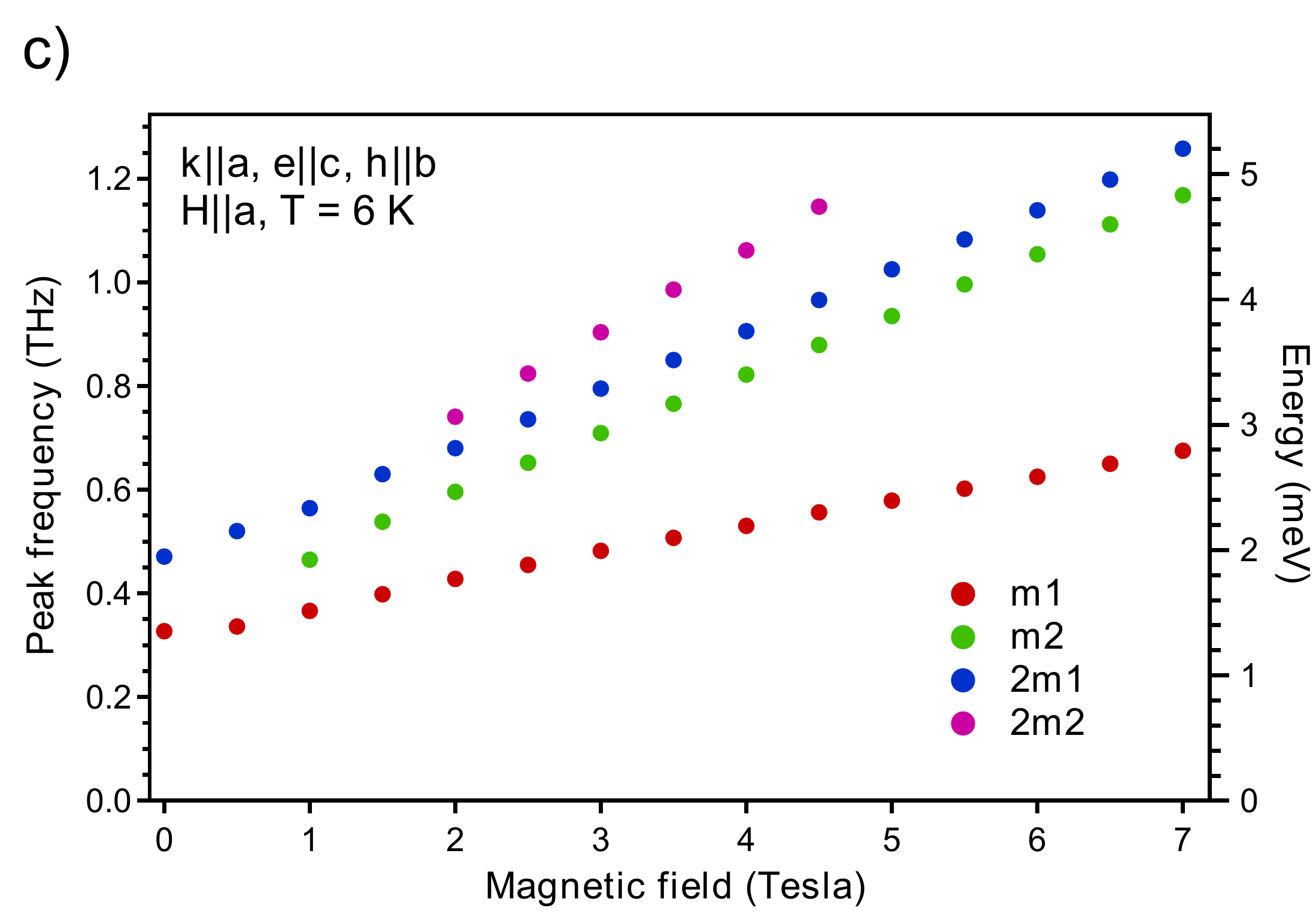}
\caption{(a) $\mathbf{k}\|c$, $\mathbf{e}\|a$, $\mathbf{h}\|b$ absorption above the incommensurate ordering temperature, 2.95 K, parallels the development of ferromagnetic correlations in the chains. The inset shows a five-spin flip excitation above this ordering temperature, with kinks represented by red circles.(b) Magnetic field dependence for $\mathbf{k}\|a$, $\mathbf{e}\|c$, $\mathbf{h}\|b$ at 6 K. (c) Peak positions for $\mathbf{k}\|a$, $\mathbf{e}\|c$, $\mathbf{h}\|b$ at 6 K. $H$ denotes the direction of the DC magnetic field.}
\label{Fig2}
\end{figure}

In Fig. \ref{Fig2}a we show the absorption spectra (wavevector $\mathbf{k} \| c$, and ac electric and magnetic fields $\mathbf{e} \| a$ and $\mathbf{h}\|b$) down to temperatures just above the 2.95 K transition to the incommensurate ordered state.  In this orientation, three features develop as the temperature is lowered: a low energy peak at 300 GHz, a higher energy peak at 500 GHz, and a broad background excitation we call the continuum. We label the 300 GHz peak the $m$ excitation and the 500 GHz peak the $2m$ excitation for reasons explained below. To determine whether the observed excitations are electric or magnetic dipole active, a full polarization dependence study was performed on three samples in a total of 6 orientations, the results of which are detailed in Table \ref{pol_table}. The observed excitations correlate with the ac magnetic field direction, and are therefore magnetic dipole active.  As Fig. \ref{Fig2}a shows, the absorption strength of these transitions increases with decreasing temperature.  This temperature dependence is consistent with the onset of ferromagnetic correlations in the chains along the c-axis near 25 K \cite{Hanawa1994}.

To understand the origin of the excitations, a magnetic field dependence was performed at 6 K, above all interchain ordering transitions, but well below the temperature where strong ferromagnetic correlations develop along chains. The magnetic field was applied along the $a$-axis, with the terahertz waveform oriented with $\mathbf{k}\|a$, $\mathbf{e}\|c$, and $\mathbf{h}\|b$, the Faraday geometry. In this configuration, the $m$, $2m$, and continuum excitations are all observed at zero field.  As Figs. \ref{Fig2}b and \ref{Fig2}c show, the $m$ and $2m$ peaks split, as predicted by the theory of Torrance and Tinkham for spin chains in a pure longitudinal field \cite{Tinkham1969a}. The slope of the lines is proportional to the total spin in the excited state. In the present case their zero field intercept reflects the number of kink pairs in the excited state.  The $m$ peak splits into a 1-spin flip excitation labeled $m1$, and a 2-spin flip excitation $m2$. The $2m$ peak splits into a $2m1$ excitation with the same slope as $m2$ (e.g. a 2-spin flip state) and a 3-spin flip $2m2$ excitation. From the zero field intercepts of these excitations \cite{Tinkham1969a}, we assign $m1$ and $m2$ as 2 kink excitations and the $2m1$ and $2m2$ as 4 kink excitations.

\begin{table}
\setlength{\tabcolsep}{10pt}
\caption{Polarization dependence of the absorption peaks at T = 3 K. An x indicates the observed presence of the feature for the particular THz polarization condition.}
\begin{tabular}{c c c c c}
\hline\hline
Polarization & m & 2m & continuum \\
\hline
$h\|a, e\|b$ &  & x & x \\ \hline
$h\|a, e\|c$ &  & x & x \\ \hline
$h\|b, e\|a$ & x & x & x \\ \hline
$h\|b, e\|c$ & x & x & x \\ \hline
$h\|c, e\|a$ & x &  &  \\ \hline
$h\|c, e\|b$ & x &  &  \\ \hline
\end{tabular}
\label{pol_table}
\end{table}

In the commensurate state below 1.97 K, the spin flip excitations should resolve into a series of bound state excitations as the effective longitudinal field from interchain interactions introduces an effective attraction between the kinks. The energy between these subdivided excitations is expected to be on the order of 0.1 meV (24 GHz).  In TDTS the total time scanned determines the spectral resolution.  In further measurements the delay stage was scanned 9 mm ($\sim$60 ps), giving an energy resolution of 0.07 meV ($17$ GHz). However, increasing the measured terahertz pulse duration complicates the analysis, as multiple reflections in the crystal are now observed (Fig. \ref{Fig1}b). These give pronounced Fabry-Perot resonance peaks at frequencies where half integral wavelengths of light fit inside the sample.  These resonance peaks could not be numerically removed sufficiently for the magnetic peak structure to be discerned. We instead use a new technique to extract high resolution spectra of magnetic excitations obscured by Fabry-Perot oscillations. First, the low temperature spectra are referenced to high resolution spectra at 3.0 K just above the inter-chain ordering temperature.  Assuming a constant index of refraction between these two temperatures, the Fabry-Perot resonances cancel. By further referencing to a temperature above the ferromagnetic chain ordering, the absorption due to magnetic transitions alone can be isolated. This analysis is described in detail in the SI. 

\begin{figure}[ht]
\includegraphics[width=.9\columnwidth]{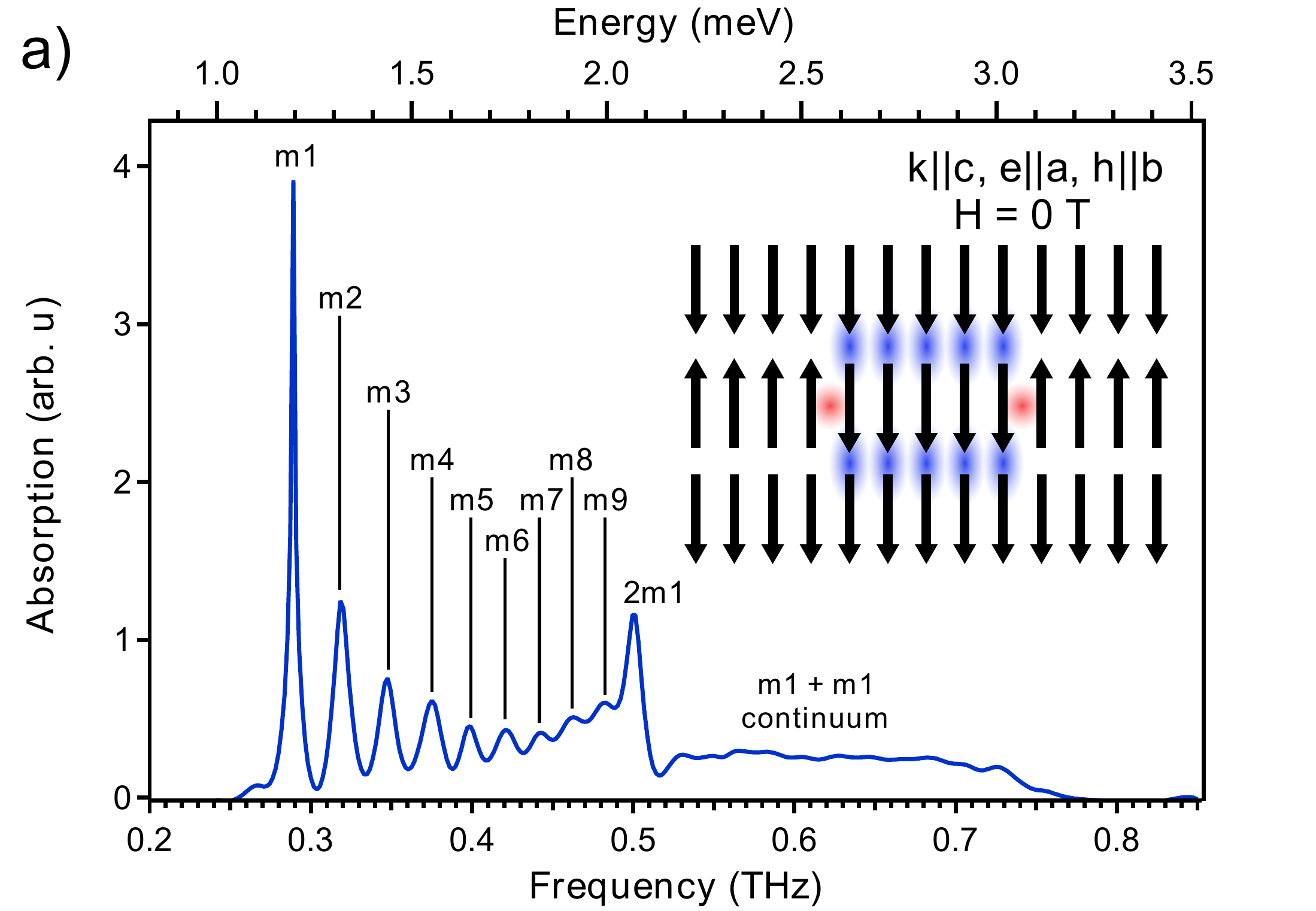}
\includegraphics[width=.9\columnwidth]{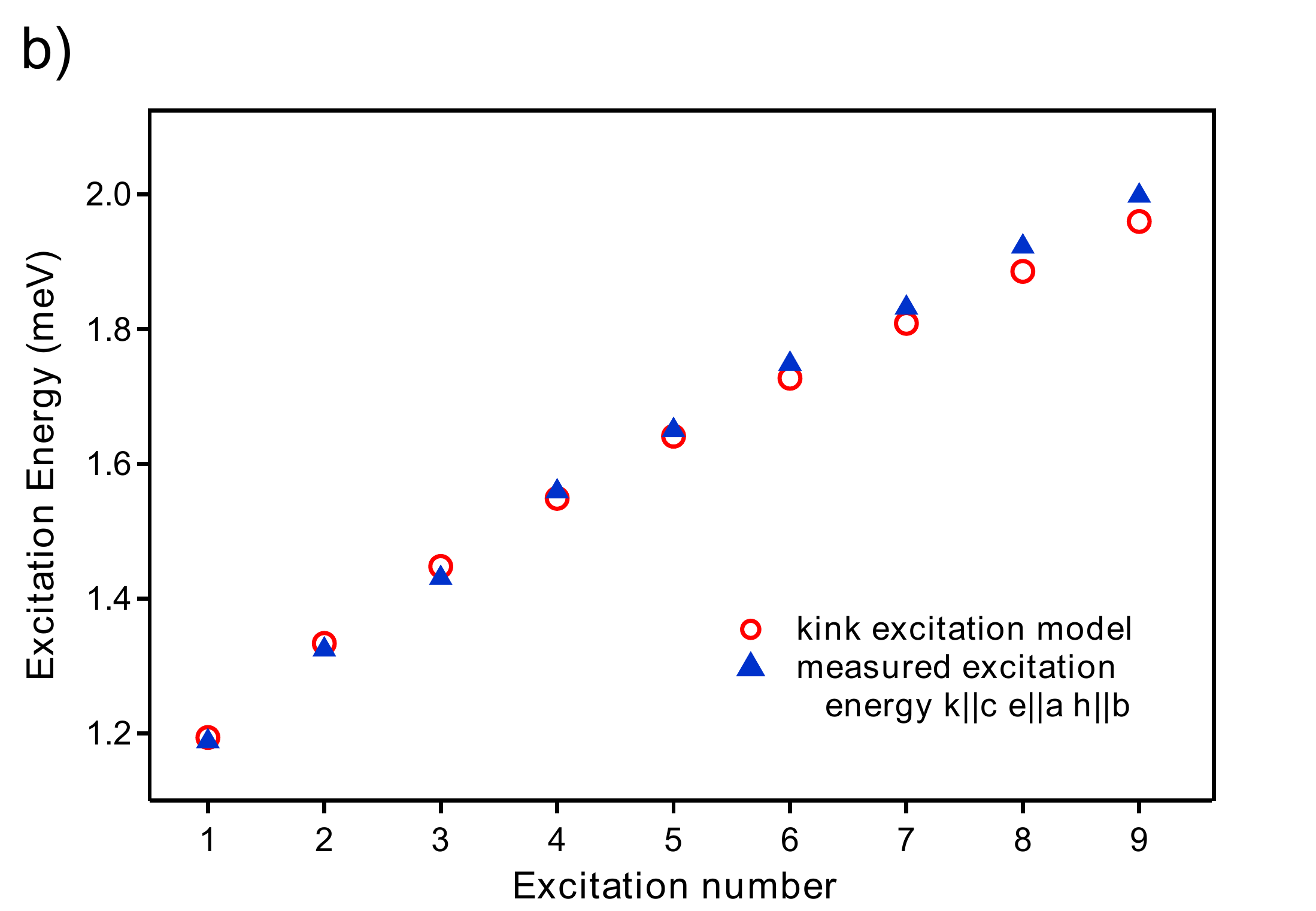}
\includegraphics[width=.9\columnwidth]{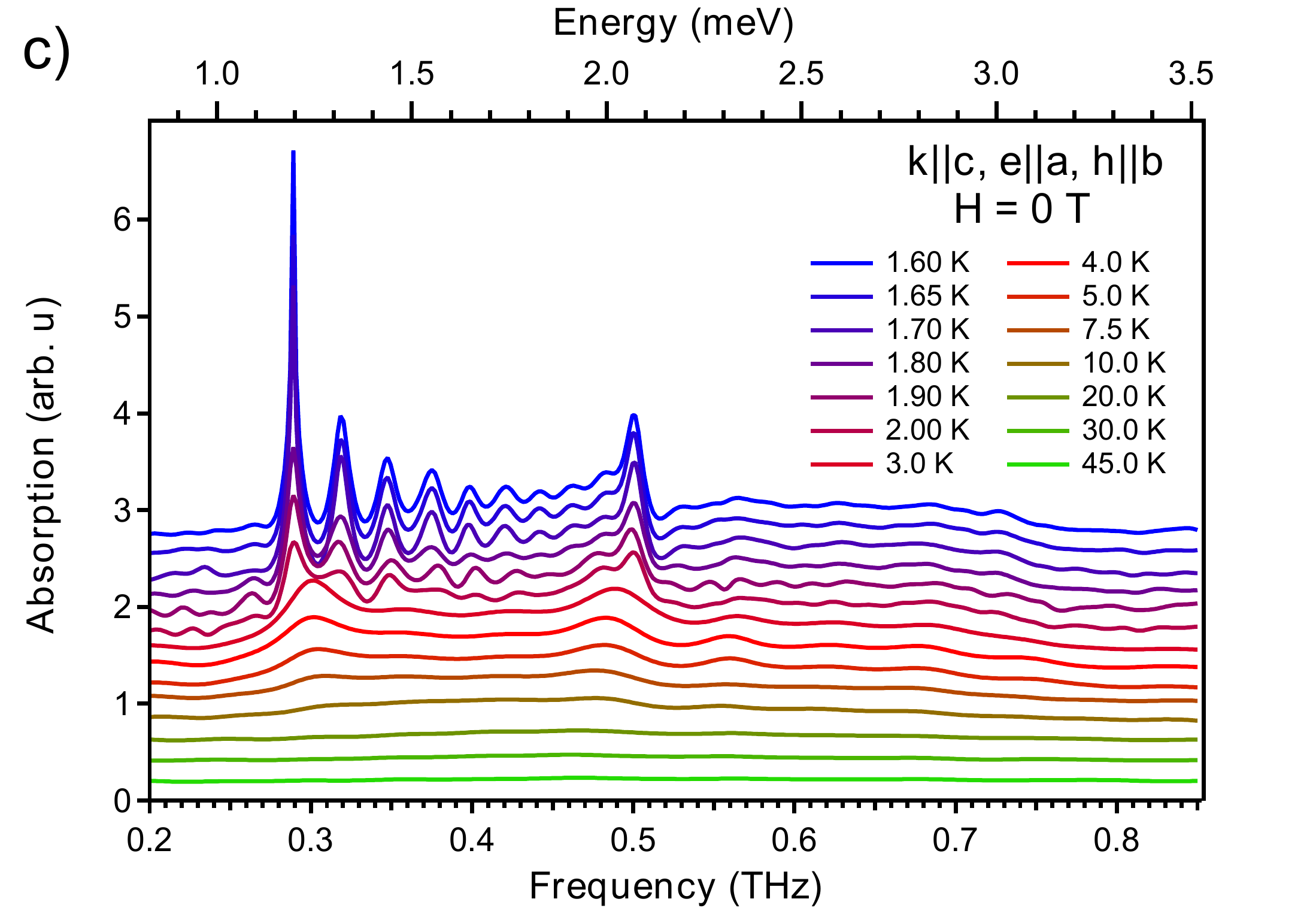}
\caption{(a) 1.6 K high resolution scan showing the hierarchy of excitations ($m1$-$m9$), the 2m1 excitation, and the continuum. The exact intensity and lineshape of the $m1$ peak is difficult to determine, as at this frequency nearly all of the light was absorbed. (b) Comparison of the energies predicted by Eq. \ref{sps} and the energies in (a). Error bars are within the width of the markers. (c) High resolution temperature dependence, showing that the kink bound states disappear as the interchain ordering is lost at $\sim3$ K.}
\label{Fig3}
\end{figure}

Fig. \ref{Fig3}a shows the result of this normalization for the high resolution spectrum at 1.6 K.  The broad absorption seen at higher temperatures subdivides into a dramatic structure of peaks.  At the lowest energies, a series of nine peaks ($m1$-$m9$) is observed, with intensity that decreases with increasing frequency.   This series terminates with a more prominent peak at 500 GHz and is then followed by a broad continuum.   As shown in Fig. \ref{Fig3}b  the energies of all nine of the lowest excitations can be described exceedingly well by the energies of the linearly confined kink model in Eq. \ref{sps}.  Our five lowest peaks have the same energies as the peaks seen in \cite{Coldea2010} at zero external field.   We identify the broad feature as the onset of the two bound state continuum at an energy which is somewhat below the energy of $m1+m1$.  The fact that no kink bound states are observed above the threshold shows the utility of the quark confinement analogy.   Free kinks (quarks) are impossible because above a threshold energy of twice the lowest bound state energy $m1$, it is energetically more efficient to excite two kink bound states $m1+m1$ (2 pairs of quarks). As discussed below, we believe the large peak ($2m1$) below the onset of the continuum is a bound state of two $m1$ kink bound states on adjacent chains.  We are able to resolve additional excitations beyond those observed by Coldea $et$ $al.$  \cite{Coldea2010} due to the high signal to noise of TDTS and its relative weighting of the absorption spectra over $\chi(\omega)$ by a factor of $\omega$. This highlights the utility of TDTS as a complementary high resolution technique for magnetic systems.

The temperature dependence of these excitations is shown in Fig. \ref{Fig3}c. At temperatures well below the ordering at 1.97 K, the excitations are well defined. As the temperature increases, they become less prominent.  As the ordering changes from commensurate to incommensurate AF near 2 K, the peaks become barely resolvable.  Above the incommensurate ordering at 3K, we see that they have reduced to their higher temperature behavior.  We see that the series of sharp peaks ($m1-m9$) evolves out of the high temperature $m$ peak and the $2m1$ peak evolves out of the $2m$ peak. 

Finally, we look at the $2m1$ peak at 500 GHz, which is associated with 4 kinks, i.e. two separate spin clusters. The energy of the peak is well below what one would expect for two isolated $m1$ excitations, which would appear at 580 GHz for the observed $m1$ excitation frequency of 290 GHz. Instead, the $2m1$ excitation appears near 500 GHz. This, and its sharpness, implies that it is associated with a bound state, as the two particle excitations give rise to the continuum.  We note that upon close inspection this feature can be seen in the data of Coldea $et$ $al.$  \cite{Coldea2010} where it appears as a dispersionless mode.  The most obvious source for binding between $m1$ excitations would come from two single spin flip excitations on adjacent chains.   A simple classical model for paired bound states with no relative motion, that is based on a modified version of Eq. \ref{sps} where one pays an energy cost $4m_0$, but only half the interaction energy per kink pair due to the fact that the spin flips are on adjacent chains, gives a rough estimate for the $2m1$ frequency of 532 GHz.   


To further investigate the $2m1$ excitation, numerical calculations were performed that predict the existence of a bound state between $m1$ excitations on adjacent chains in a manner proposed above. With the accepted parameters for CoNb$_2$O$_6$ the energy of this bound state was calculated to be 564 GHz, lying below the calculated bottom of the $m1+m1$ continuum at $2 \times286$ GHz $=572$ GHz. While the quantitative agreement is not exact, qualitatively the existence of a bound state below the continuum is confirmed.  Moreover, calculated dispersion curves show that the $2m1$ mode has a significantly reduced dispersion compared to the $m1$ state, although again the calculated dispersion is not as flat as that observed by Coldea $et$ $al.$.   Consistent with observation, the spectral weight in this novel excitation is predicted to be appreciable and of order the weight in the $m2$ peak. See the SI for further details.

We have reported the observation of 9 kink bound states in the one dimensional spin chain CoNb$_2$O$_6$ by high resolution time-domain terahertz spectroscopy.  Their energies can be modeled exceedingly well by the Airy function solutions to a 1D Schr{\"o}dinger equation in a linear confining potential.  This sequence of bound states terminates at a threshold energy two times the lowest bound state energy.  Above this energy scale we observe a broad feature consistent with the onset of the two particle continuum.  At energies just below this threshold we observe a prominent excitation at an energy somewhat less than two times the lowest bound state.   We interpret this feature as resulting from a novel bound state of bound states on neighboring chains.  These results highlight the complementary role that terahertz spectroscopy can play to neutron studies of magnetic systems.

We would like to thank C. Broholm,  I. Cabrera, J. Deisenhofer, J. Kj\"all, J. Moore, K. Ross, and M. Mourigal for helpful discussions.  The THz measurements and instrumentation development was funded by the Gordon and Betty Moore Foundation through Grant GBMF2628 to NPA.  The crystal growth and theoretical work was funded by the DOE-BES through DE-FG02-08ER46544.

\bibliography{QuantumMag}

\newpage
\begin{widetext}
\section{Supplementary materials for ``A hierarchy of bound states in the 1D ferromagnetic Ising chain CoNb$_2$O$_6$ investigated by high resolution time-domain terahertz spectroscop"}

\subsection{Terahertz technique}

The time domain terahertz spectroscopy setup used in this work is shown in Fig. S1. A KM Labs Ti:Sapph modelocked laser is used to generate $\sim$60 femtosecond duration NIR pulses centered at 820 nm with an 80 MHz repetition rate. A beam splitter divides each pulse into two different paths. The first part of the pulse is used to excite a photoconductive Auston-switch antenna. This antenna is voltage biased, which causes charge to accelerate across the small antenna gap, emitting a few picosecond THz pulse. This THz pulse is collimated and focused onto the sample with a silicon lens and a pair of off-axis parabolic mirrors (OAPs). The transmitted pulse is then focused with a second pair of OAPs onto a detector antenna.  The second part of the NIR pulse arrives at the receiver antenna at the same time as the THz pulse and photoexcites charge carriers in the antenna gap. The electric field of the THz pulse drives the photoexcited charge across the antenna gap, producing a current. The current is converted to a voltage using an SRS 570 transimpedance amplifier and measured by an SRS 830 lockin amplifier.  A delay stage is used to vary the arrival time of the THz pulse at the detector antenna and map out the full THz electric field as a function of time. A mechanical chopper is used to modulate the NIR light incident on the emitter at $\sim800$ Hz. The sample is mounted in a helium-vapor cryostat capable of temperatures ranging from $1.6-300$ K.

Once the full time domain THz electric field is collected, it is Fourier transformed to obtain the full complex transmission of the sample.  As both the amplitude and phase of the THz electric field are collected, the full complex transmission coefficient can be obtained without use of Kramers-Kronig analysis.

\subsection{THz Analysis}

As described in the main text, in order to achieve sufficiently high resolution to discern the individual $m1-m9$ excitations, the time domain electric field is collected for 60 ps.  However, this long collection time introduces difficulties, as the spectroscopy is being performed on a 600 $\mu$m thick single crystal of CoNb$_2$O$_6$ with an index of refraction of $\sim5.4$. With these parameters, reflections of the THz pulse are observed in the time domain spaced by approximately 22 picoseconds.  These reflected pulses in the time domain manifest themselves as Fabry-Perot oscillations in the frequency spectrum, with an amplitude large enough to obscure the fine magnetic structure. In principle sufficient knowledge of the frequency dependent index of refraction should make it possible to numerically remove these oscillations, however, in practice the required level of precision needed in the index of refraction is too high. Instead, we developed a method of data analysis that exploits the fact that the index of refraction of the material changes fairly slowly with temperature (aside from the absorption due to magnetic excitations), and therefore referencing between scans with similar temperatures will remove the oscillations.

\begin{figure}[tb]
\includegraphics[width=.9\columnwidth]{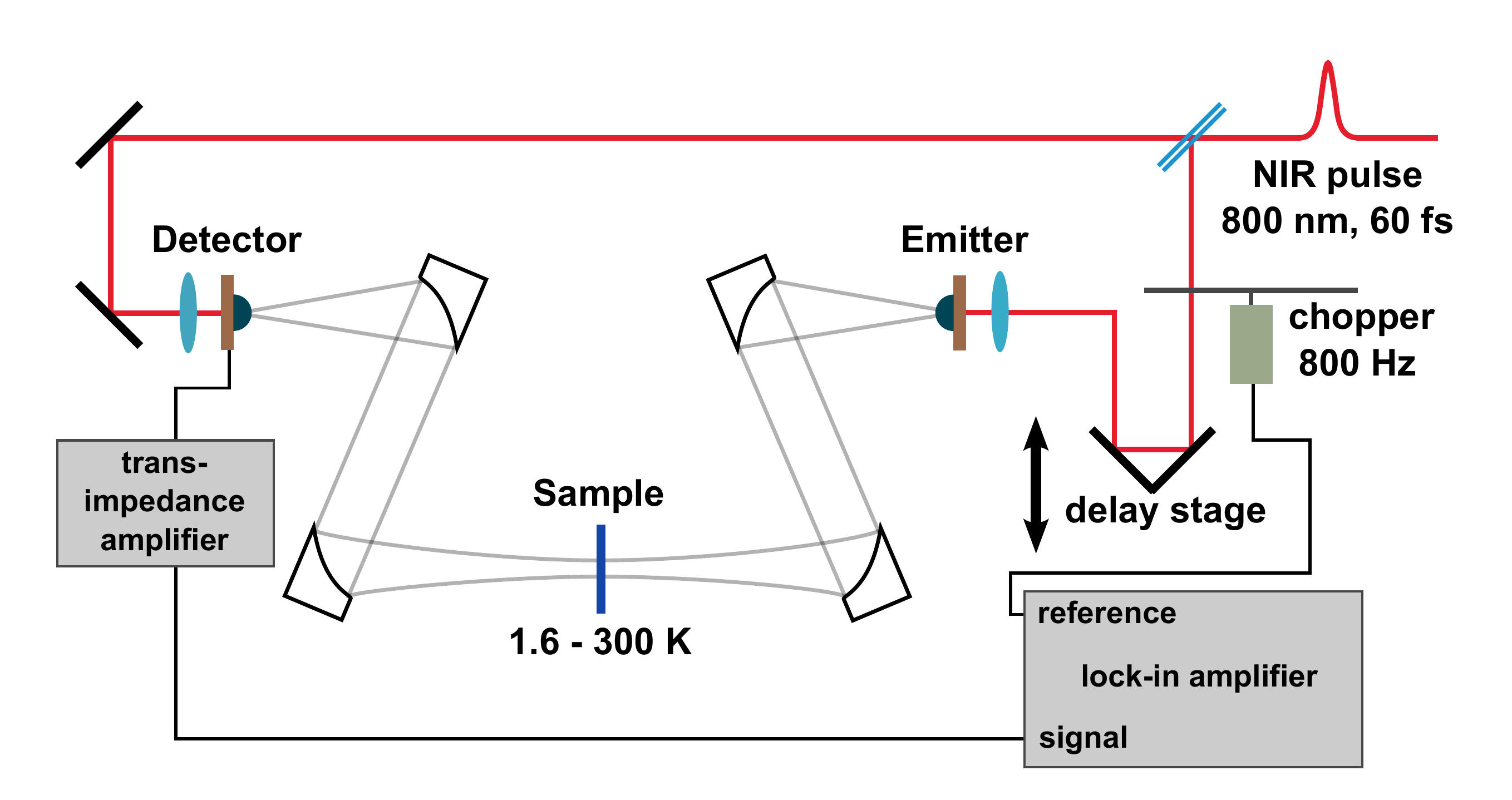}
\begin{center}
Fig. S1: Time domain terahertz spectroscopy transmission experimental setup.
\end{center}
\label{FigS1}
\end{figure}

The measured spectral amplitude of the THz pulse is given by:

\begin{eqnarray}
S_{measured}\left(\omega,T,\tau\right) & = & S\left(\omega,\tau\right)\cdot t_{magnetic}\left(\omega,T,\tau\right)\nonumber \\
 &  & \cdot t_{background}\left(\omega,T,\tau\right)\cdot t_{Fabry-Perot}\left(\omega,T,\tau\right)
\end{eqnarray}

\noindent where $\omega$ is the terahertz frequency, $T$ is the temperature
where the measurement is made, and $\tau$ is the time length of the collected scan. This time $\tau$ directly determines the frequency resolution of the spectrum ($\Delta\omega=2\pi/\tau$). Additionally, we can choose different $\tau$ to eliminate the Fabry-Perot oscillations at the expense of spectral resolution, which we will use later. We denote high resolution scans by the long pulse duration $\tau_L$ and low resolution scans by the shorter THz pulse duration $\tau_S$. $S\left(\omega,\tau\right)$ is the spectrum of the initial waveform incident on the CoNb$_{2}$O$_{6}$ crystal, $t_{magnetic}$ is the transmission change due to the magnetic excitations, $t_{background}$ is the change in transmission due to any non-magnetic sources (here mainly cause by reflection at the surface of the crystal and essentially $T$ independent for the temperatures used here), and $t_{Fabry-Perot}$ characterizes the Fabry-Perot oscillations caused by the reflections of the pulse within the crystal.

Our goal is to resolve the magnetic excitations in the system $m1-m9$ at temperature $T_0$ below the antiferromagnetic interchain ordering temperature 1.97 K. The first step in the analysis is to divide the high resolution spectrum at $T_{0}$ by a spectrum at a nearby temperature $T_{R1}$, to remove the Fabry-Perot oscillations.

\begin{eqnarray}
\frac{S\left(\omega,T_{0},\tau_{L}\right)}{S\left(\omega,T_{R1},\tau_{L}\right)} & = & \frac{S\left(\omega,\tau_{L}\right)\cdot t_{magnetic}\left(\omega,T_{0},\tau_{L}\right)}{S\left(\omega,\tau_{L}\right)\cdot t_{magnetic}\left(\omega,T_{R1},\tau_{L}\right)}\nonumber \\
& & \cdot\frac{t_{background}\left(\omega,T_{0},\tau_{L}\right)\cdot t_{Fabry-Perot}\left(\omega,T_{0},\tau_{L}\right)}{t_{background}\left(\omega,T_{R1},\tau_{L}\right)\cdot t_{Fabry-Perot}\left(\omega,T_{R1},\tau_{L}\right)}
\end{eqnarray}

As the background reflection is expected to change very slowly with temperature, the two $t_{background}$ terms will cancel. If the magnetic absorption is weak and the change in the index of refraction is small between the temperatures $T_0$ and $T_{R1}$, the two $t_{Fabry-Perot}$ terms cancel as well, leaving us with

\begin{equation}
\frac{S\left(\omega,T_{0},\tau_{L}\right)}{S\left(\omega,T_{R1},\tau_{L}\right)}=\frac{t_{magnetic}\left(\omega,T_{0},\tau_{L}\right)}{t_{magnetic}\left(\omega,T_{R1},\tau_{L}\right)}
\end{equation}

This brings us closer to isolating the high resolution magnetic response at $T_0$.  However, the magnetic absorption at $T_{R1}$ is still significant, as in our analysis we used $T_{R1}=3.0$ K to cancel the Fabry-Perot oscillations as fully as possible. In order to isolate the high resolution transmission at the low temperature $T_0$, we need to account for the magnetic absorption signal at the higher temperature $T_{R1}$. We note that as shown in Fig. 2a in the main text, absorption above the ordering temperature is intrinsically broad, and eliminating reflections by only collecting for $\tau_S$ will still allow the absorption at $T_{R1}$ to be fully resolved.  To remove $t_{magnetic}$ at $T_{R1}$, we multiply by the ratio of the low resolution ($\tau_S$) spectra at $T_{R1}$ and a new higher reference temperature ($T_{R2}$) that is well above the onset of ferromagnetic correlations that begin near $25$ K. The low resolution scans of length $\tau_S$ avoid Fabry-Perot oscillations in the spectra.  Here we use $T_{R2} = 60$ K.

\begin{eqnarray}
\frac{S\left(\omega,T_{0},\tau_{L}\right)}{S\left(\omega,T_{R1},\tau_{L}\right)}\cdot\frac{S\left(\omega,T_{R1},\tau_{S}\right)}{S\left(\omega,T_{R2},\tau_{S}\right)} & = &\frac{t_{magnetic}\left(\omega,T_{0},\tau_{L}\right)}{t_{magnetic}\left(\omega,T_{R1},\tau_{L}\right)}\cdot\frac{S\left(\omega,\tau_{S}\right)\cdot t_{magnetic}\left(\omega,T_{R1},\tau_{S}\right)}{S\left(\omega,\tau_{S}\right)\cdot t_{magnetic}\left(\omega,T_{R2},\tau_{S}\right)} \nonumber \\
& & \cdot\frac{t_{background}\left(\omega,T_{R1},\tau_{S}\right)\cdot t_{Fabry-Perot}\left(\omega,T_{R1},\tau_{S}\right)}{t_{background}\left(\omega,T_{R2},\tau_{S}\right)\cdot t_{Fabry-Perot}\left(\omega,T_{R2},\tau_{S}\right)}
\end{eqnarray}

We assume again the the background absorption at $T_{R1}$ and $T_{R2}$ are identical, and thus cancel.  For these short time scans, $t_{Fabry-Perot}$ = 1, which leaves

\begin{equation}
\frac{S\left(\omega,T_{0},\tau_{L}\right)}{S\left(\omega,T_{R1},\tau_{L}\right)}\cdot\frac{S\left(\omega,T_{R1},\tau_{S}\right)}{S\left(\omega,T_{R2},\tau_{S}\right)}=\frac{t_{magnetic}\left(\omega,T_{0},\tau_{L}\right)}{t_{magnetic}\left(\omega,T_{R1},\tau_{L}\right)}\cdot\frac{t_{magnetic}\left(\omega,T_{R1},\tau_{S}\right)}{t_{magnetic}\left(\omega,T_{R2},\tau_{S}\right)}
\end{equation}

We specifically chose $T_{R2}$ to be above the ferromagnetic ordering temperature, so that the magnetic transmission component $t_{magnetic}\left(\omega,T_{R2},\tau_{S}\right)=1$ and we are left with

\begin{equation}
\frac{S\left(\omega,T_{0},\tau_{L}\right)}{S\left(\omega,T_{R1},\tau_{L}\right)}\cdot\frac{S\left(\omega,T_{R1},\tau_{S}\right)}{S\left(\omega,T_{R2},\tau_{S}\right)}=t_{magnetic}\left(\omega,T_{0},\tau_{L}\right)\cdot\frac{t_{magnetic}\left(\omega,T_{R1},\tau_{S}\right)}{t_{magnetic}\left(\omega,T_{R1},\tau_{L}\right)}
\end{equation}

As described earlier, above the ordering temperature the magnetic absorption becomes broad, such that the high and low resolution scans should have the same linewidth, and thus their ratio should be 1, leaving us with just the transmission due to the high resolution scan in the ordered state at $T_0$.

\begin{equation}
\frac{S\left(\omega,T_{0},\tau_{L}\right)}{S\left(\omega,T_{R1},\tau_{L}\right)}\cdot\frac{S\left(\omega,T_{R1},\tau_{S}\right)}{S\left(\omega,T_{R2},\tau_{S}\right)}=t_{magnetic}\left(\omega,T_{0},\tau_{L}\right)
\end{equation}

We have confirmed that the final high resolution magnetic absorption spectra do not depend on the precise temperatures chosen or the exact time intervals chosen for $\tau_L$ and $\tau_S$.

\subsection{Energy of the $2m1$ excitation}

\begin{figure}[tb]
\includegraphics[width=.9\columnwidth]{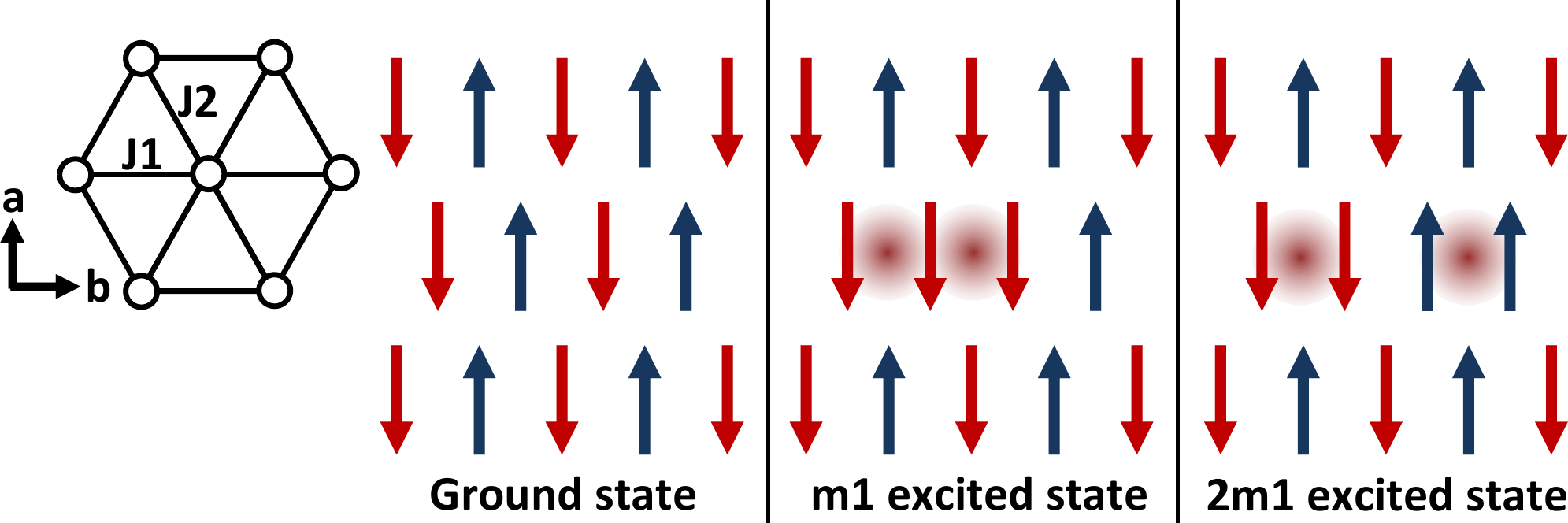}
\begin{center}
Fig. S2: Diagram of the $m1$ and $2m1$ excitations in the low temperature ordered state. The exchange interaction along the $b$-direction has a strength $J_1$, while along the diagonals in the ab-plane the interaction strength is $J_2$. Both the $m1$ and $2m1$ break two $J_1$ exchange interactions while leaving the total energy due to $J_2$ exchange unchanged.
\end{center}
\label{FigS2}
\end{figure}

The energy of a bound state excitation on a single chain ($m1-m9$) can be calculated in the confined kink model using Eq. 2 in the main text.  There, $2m_0$ is the energy cost of the ferromagnetic intrachain exchange interaction at the two domain walls. The second term, $E_{mf} = z_1 \lambda^{2/3} \left(\frac{\hbar^2}{\mu}\right)^{1/3}$, is the energy cost to break the mean field interaction caused by the interchain ordering.

To create two $m1$ excitations, 4 domain walls must be created at an energy cost of $4m_0$. If the excitations occur on the same chain, each $m1$ excitation pays the cost of the mean field interaction, for a total excitation energy of $4m_0 + 2E_{mf}$, double the energy of a single $m1$ excitation. As Fig. 3a in the main text shows, the observed energy is lower than this. However, if two $m1$ excitations are created on sites on adjacent chains in the ab-plane, the mean field energy cost is reduced from $2E_{mf}$ to $E_{mf}$, so that the total energy will be

\begin{equation}
E_{2m1} = 4m_0 + z_1 \lambda^{2/3} \left(\frac{\hbar^2}{\mu}\right)^{1/3}
\end{equation}

which is much closer to the observed $2m1$ energy. We use the values for the constants from Coldea et al., with $m_0=0.5035$ meV and $\lambda^{2/3} \left(\hbar^2/\mu\right)^{1/3}=0.0798$ meV. The mechanism of this excitation is depicted in Fig. S2.  The chains are shown below the ordering temperature of 1.97 K.  Each arrow represents a ferromagnetically aligned chain coming out of the page in the c-direction. The chains order antiferromagnetically along the b-direction. The exchange interactions along the b-direction have a strength $J_1$, while along each diagonal the exchange interaction strength is $J_2$, with both interactions being antiferromagnetic.

When a single spin flip occurs, there is no change in the total energy due to exchange along the diagonals, $J_2$. However, the exchange interaction $J_1$ is broken between the two neighboring spins in the b-direction, which costs total energy $E_{mf}$. 

When two spin flips occur simultaneously on neighboring spin chains, the $2m1$ excited state is formed. Again, the total energy of the exchange interaction due to $J_2$ remains the same, and two $J_1$ interactions are broken, again at a total cost $E_{mf}$, making the total energy of the excited state $4m_0 + E_{mf}$.

\subsection{Sample preparation}

Stoichiometric amounts of Co$_3$O$_4$ and Nb$_2$O$_5$ were thoroughly ground together by placing them in an automatic grinder for 20 minutes. Pellets of
the material were pressed and heated at 950$^\circ$ C with one intermittent
grinding. The powder was then packed, sealed into a rubber tube evacuated
using a vacuum pump, and formed into rods (typically 6 mm in diameter
and 70 mm long) using a hydraulic press under an isostatic pressure
of $7$x$10^7$ Pa. After removal from the rubber
tube, the rods were sintered in a box furnace at 1375$^\circ$ C for 8 hours in
air.

Single crystals approximately 5 mm in diameter and 30 mm in length
were grown in a four-mirror optical floating zone furnace at Johns Hopkins (Crystal
System Inc. FZ-T-4000-H-VII-VPO-PC) equipped with four 1-kW halogen
lamps as the heating source. Growths were carried out under 2 bar
O$_2$-Ar (50/50) atmosphere with a flow rate of 50 mL/min, and a zoning
rate of 2.5 mm/h, with rotation rates of 20 rpm for the growing crystal
and 10 rpm for the feed rod. Measurements were carried out on oriented
samples cut directly from the crystals using a diamond wheel.

Cut samples were polished to a finish of 3 $\mu$m and total thickness of $\sim600$ $\mu$m using diamond polishing paper and a specialized sample holder to ensure that plane parallel faces were achieved for the THz measurement. The discs were approximately 5 mm in diameter. Three different orientations were cut, each with a different crystallographic axis oriented normal to the sample face.

\subsection{The four-kink bound state}

\begin{figure}[tb]
\includegraphics[width=.96\columnwidth]{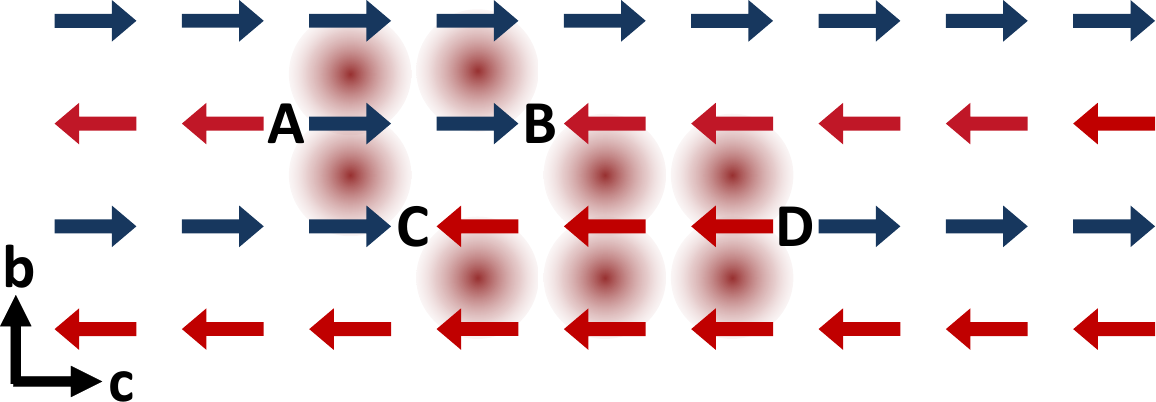}
\begin{center}
Fig. S3: A typical four-kink configuration. The four kinks where intrachain order is broken at an energy cost of $J/2$ each are labeled as A, B, C and D. Sites where interchain ordering is broken are shown by colored red clouds.
\end{center}
\label{FigS3}
\end{figure}

The observation of the new $2m1$ bound state of $m1$ excitations on adjacent chains is one of the major results of the work. To further understand the nature of this excitation, we carried out calculations of its energy and dispersion. We start with a system of four kinks, two each on two adjacent Ising chains. It is clear from Fig. S2 that only the exchange interaction $J_{1}$ contributes to kink confinement and the corresponding effective longitudinal field is $h_{z}=4J_{1}$.  Hence, while studying kink bound states, we can safely ignore the couplings between the different chains in the $a$-direction and work in just two dimensions (the $c-b$ plane). The four-kink potential then depends on which $J_{1}$ bonds are frustrated in a given configuration. One can read off this potential directly from Fig. S3, which takes the form
\begin{eqnarray*}
V(x_{A},x_{B},x_{C},x_{D})&=&4J_{1}[(x_{D}-x_{C})+(x_{B}-x_{A})] \,\,\, if \,\,\, x_{D}<x_{A}\,\,\, or \,\,\,x_{B}<x_{C}\\
                                        &=&2J_{1}[|x_{A}-x_{C}|+|x_{B}-x_{D}|] + 2J_{1}[(x_{D}-x_{C})+(x_{B}-x_{A})] \,\,\, otherwise 
\end{eqnarray*}
In addition, there is the hard-wall constraint $x_{B}>x_{A}$ and $x_{D}>x_{C}$ between the kinks on any one Ising chain. Thus the effective Hamiltonian for this four-kink system consists of two copies of the two-kink Hamiltonian which appears in the supplementary material to Coldea et. al. [5] with $V$ replacing the simple linear two-kink attractive potential. It is interesting to note that the two spin-clusters (on the two adjacent $b$-chains) attract each other only when there is an overlap between them. Otherwise, they are essentially free. We found out that this inter-cluster potential, despite having a strictly finite range, accomodates at least one bound state. 

To see this, we first write the four-kink Hamiltonian in a basis labelled by the centres-of-mass (COMs) of the two clusters $i$ and $j$ and their lengths $m>0$ and $n>0$.
\begin{eqnarray*}
H|i,m;j,n>=(2J+V(i,j,m,n))|i,m;j,n>-\alpha[|i+\frac{1}{2},m+1;j,n>+|i+\frac{1}{2},m-1;j,n>\\+|i-\frac{1}{2},m+1;j,n>+|i-\frac{1}{2},m-1;j,n>+|i,m;j+\frac{1}{2},n+1>+|i,m;j+\frac{1}{2},n-1>\\+|i,m;j-\frac{1}{2},n+1>+|i,m;j-\frac{1}{2},n-1>]-\beta_{0}\delta_{m,1}[|i-1,m;j,n>+|i+1,m;j,n>]\\-\beta_{0}\delta_{n,1}[|i,m;j-1,n>+|i,m;j+1,n>]+\beta_{1}\delta_{m,1}|i,m;j,n>+\beta_{1}\delta_{n,1}|i,m;j,n>
\end{eqnarray*}
Here $J$ is the energy cost to create a pair of kinks on one Ising chain, $\alpha$ is the effective transverse field causing the kinks to hop, $\beta_{0}$ is the kinetic energy gain for two nearest neighbor kinks hopping together and $\beta_{1}$ is the extra energy cost of two adjacent kinks (i.e. a single spin flip) [5]. 
One can now transform to a more natural set of coordinates: the COM of the whole system $u=(i+j)/2$ and the relative separation of the two clusters $v=(i-j)$. The Hamiltonian has a discrete translational symmetry with respect to $u$ and hence can be partially diagonalized by Fourer transforming from the basis $\{|u,v;m,n>\}$ to the basis $\{|k,v;m,n>\}$. The problem now becomes a 3D lattice model, ${v,m,n}$ being the lattice coordinates, for every wavevector $k$ within the first Brillouin zone$\,[-\pi,\pi]$.

\begin{figure}[ht]
\includegraphics[width=0.49\columnwidth]{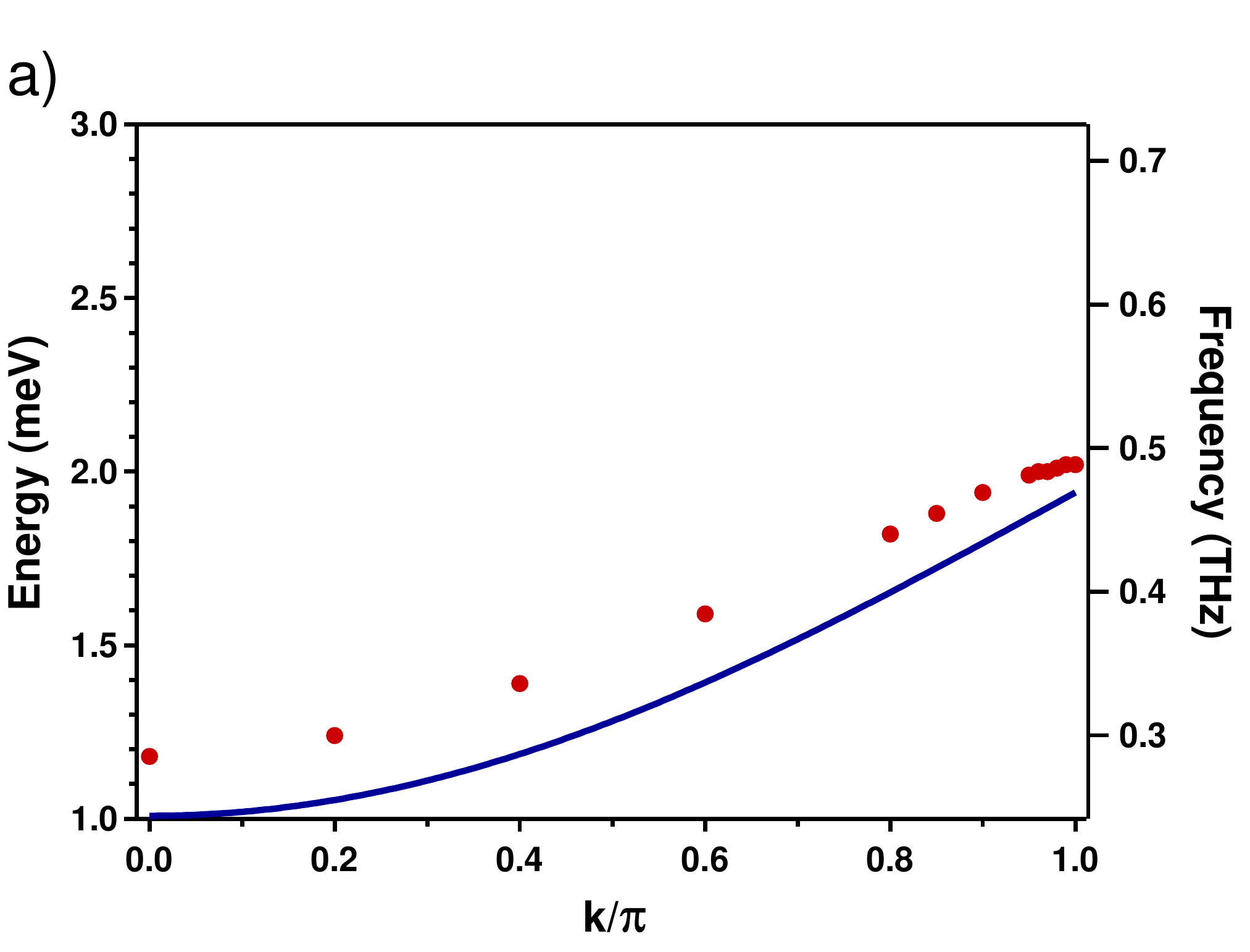}
\hskip 0.2cm
\includegraphics[width=0.49\columnwidth]{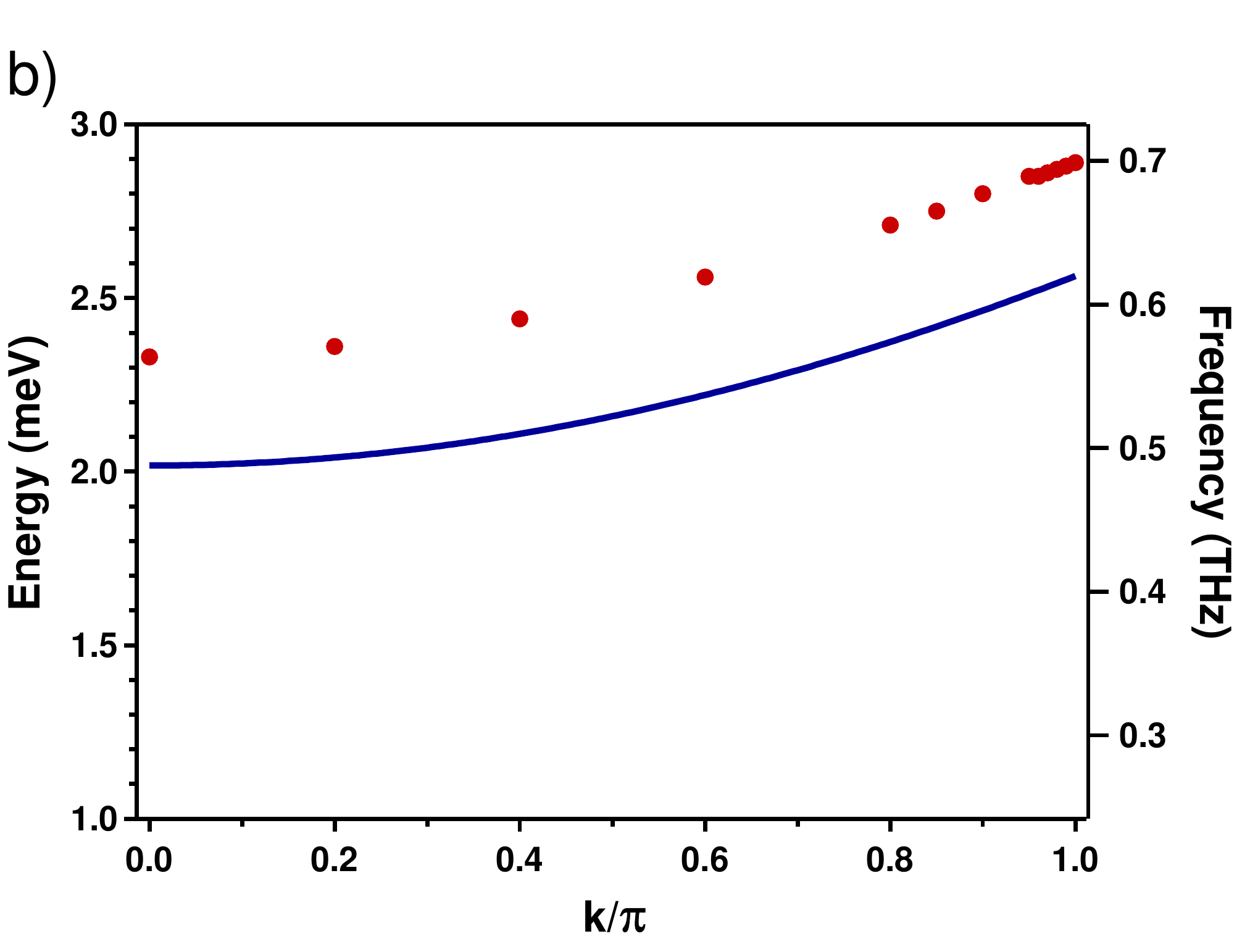}
\includegraphics[width=0.46\columnwidth]{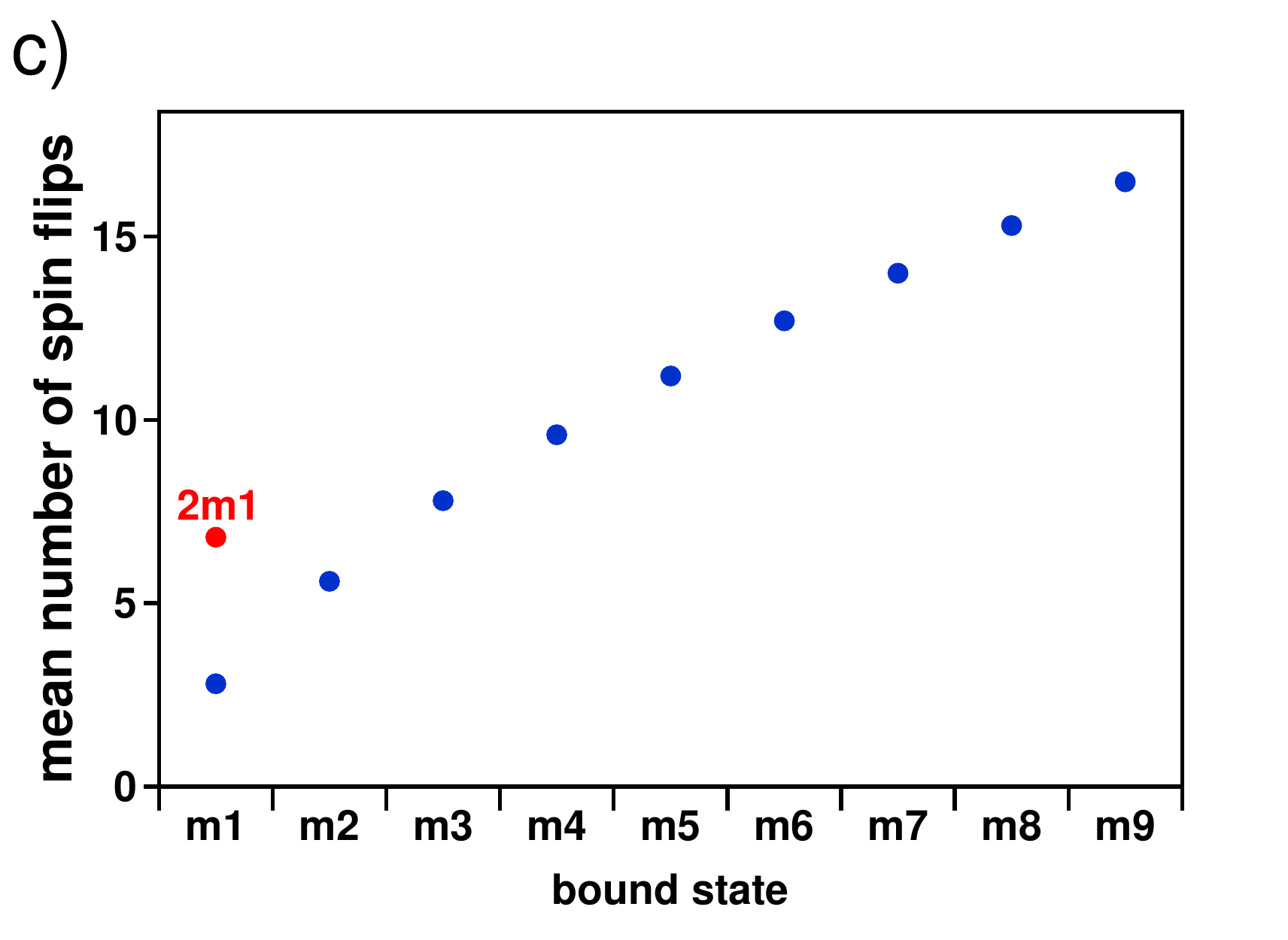}
\includegraphics[width=0.49\columnwidth]{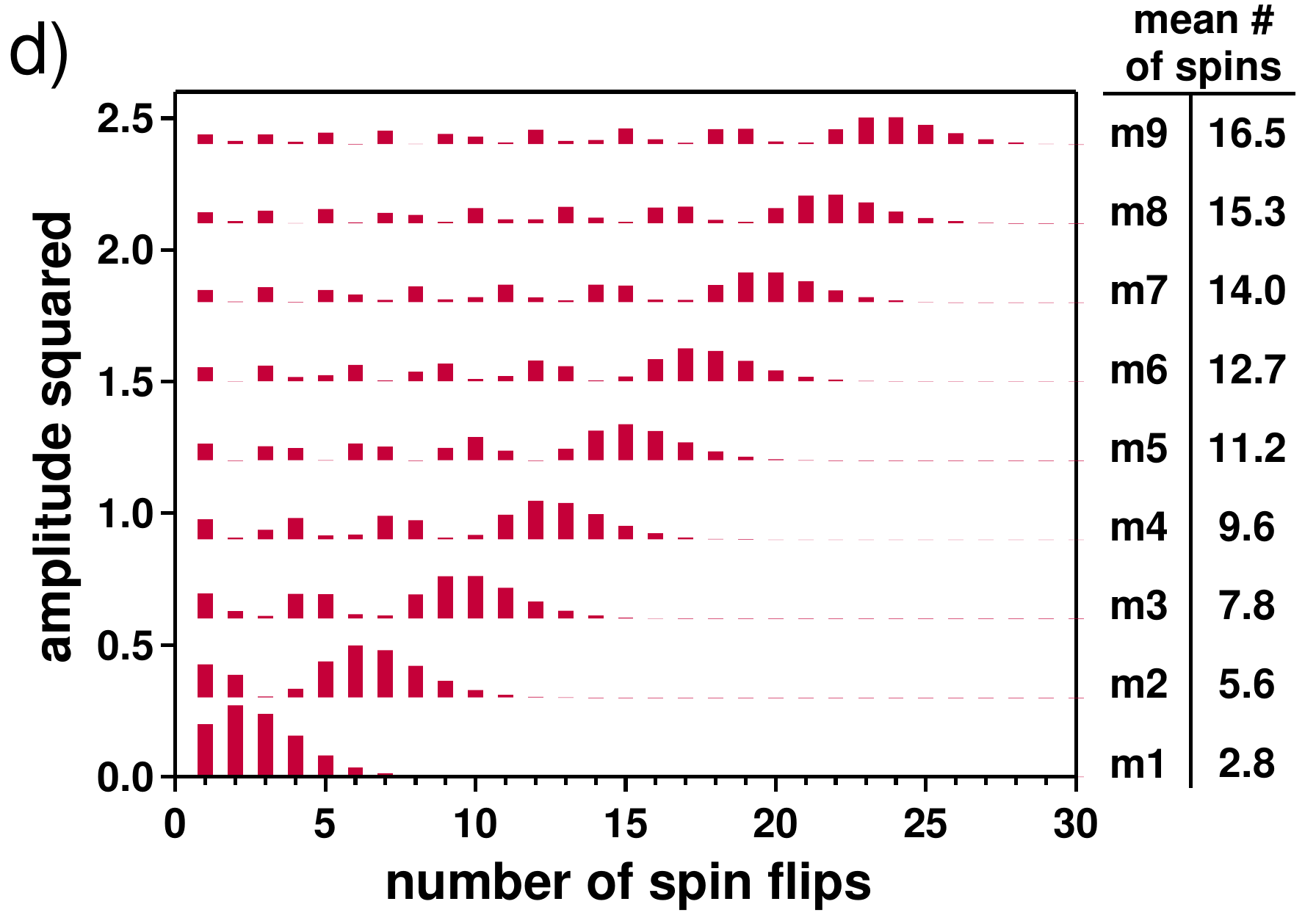}
\begin{center}
Fig.S4: Dispersion of the ground state of the (a) two-kink system (b) four-kink system  (red points). The dispersions are flat near the zone center and sharp near the edge, which match the features of the bottoms of the corresponding continua at zero longitudinal field (blue lines). (c) The weighted mean number of spin flips in the m1 - m9 and 2m1 states. (d) Squared amplitudes for the wavefunctions for m1 - m9 states in the spin flip basis.  Numerical values of the weighted mean number of spin flips shown in (c) are listed in the table.
\end{center}
\label{Fig.S4}
\end{figure}

The spectrum of the four-kink system definitely has a continuum because two non overlapping spin-clusters are essentially free. We tried to investigate if the spectrum has bound states in addition to this continuum. To do this, we first studied a two-kink system on one Ising chain which can easily be solved numerically. The  two kinks in the ground state of this system are well localized within $L \sim 10$ lattice units from each other. This implies that if there are four-kink bound states, the values of $m$ and $n$ in the lowest-lying state will be localized approximately within the same distance $L$. In that case, the range of the inter-cluster potential is $v\in[-L,L]$ and outside this range the four-kink wave-function should fall off exponentially.

Thus, if there is a bound state, the wave function will be localized within $[-aL,aL]$ in the $v$-direction for some $a$. Because the $v$ coordinate can assume half integer values, this means that we now need to find the lowest eigenvalue and the corresponding eigenstate of our reduced model within an $L \times L \times 4aL$ lattice with hard wall boundary conditions at the six faces. We can then vary $L$ and $a$ to check convergence of the energy. For $L=10$ and $a=2$ this reduces to the diagonalization problem of a $8000 \times 8000$ sparse matrix which can be easily done numerically. 

As a starting point, we took $a=2$. At $k=0$ the ground state energy converges to $2.33 \, \mathrm{meV}=564 \, \mathrm{GHz}$ for $L=8$ and stays constant (up to the second decimal place) for $L=8,9,10,11$. Increasing $a$ to $3$ does not change the energy either (up to the fifth decimal place). This indicates that we are actually seeing a stable bound state because $E_{bound}(k=0)$ is less than the bottom of the two-cluster continuum at $572 \, \mathrm{GHz}$ that we find from our calculations. 

The calculation was repeated for finite wavevectors and the dispersion of this bound state $E(k)$ in the first Brillouin zone was obtained. A plot of this dispersion is shown in Fig. S4. (Due to symmetry of the Hamiltonian, $E(k)=E(-k)$.) A plot of the two-kink ground state dispersion is also shown for comparison. It can be seen that the four-kink state is less dispersive than the two-kink state. This is expected because in the absence of any interaction between kinks, the spectrum forms a continuum. If a weak interaction is then turned on, the ground state dispersion is expected to track the bottom envelope of this continuum. It is easily seen that the bottom of the four kink coninuum is indeed flatter than the two kink one. In addition, these continua are flat at $k=0$ and have sharp edges at $k=\pi$. These features can also be seen in the bound state dispersions.

If we strictly confine ourselves to the four-kink Hilbert space, the spectral weight of the above bound state (which depends on its overlap with the single spin flip state) vanishes. But in the real material a kink pair can be created/annihilated on an Ising chain by the transverse field term in the Hamiltonian. This tunnelling mechanism between the the above state and the two-kink eigenstates endows it with a finite spectral weight which can be calculated in first-order perturbation theory. The spectral weight thus obtained for the $2m1$ excitation is $0.17$ which lies roughly halfway between the (theoretical) $m1$ and $m2$ spectral weights (0.20 and 0.13 respectively). This theoretical spectral weight is greater than what is observed experimentally from the absorption (taking into account the relative weighting of $\chi(\omega)$ by $\omega$). However, it qualitatively supports the interpretation of the $2m1$ peak as a separate excitation from the $m1$-$m9$ series, not a tenth $m10$ bound state. The theoretically predicted spectral weights for the $m1$, $m2$, etc. excitations decrease monotonically, with the $m9$ and $m10$ spectral weights being only 0.037 and 0.033, respectively. The significantly increased spectral weight of the $2m1$ excitation compared to the $m9$ excitation therefore supports the interpretation of the $2m1$ excitation as a bound state of $m1$ bound states on adjacent chains.
\newpage
\end{widetext}
\end{document}